%% file: main.tex
\documentclass[runningheads]{llncs}
\usepackage{cite}
\usepackage[T1]{fontenc}
\usepackage{graphicx}
\usepackage{xcolor}
\usepackage{booktabs}
\usepackage{dokeeffe-macros}
\usepackage{ollie-macros}
\usepackage{tikz}
\usepackage{multirow}
\usepackage{amssymb}
\usepackage{afterpage}
\usepackage{dirtree}

\usetikzlibrary{shapes,arrows,positioning,fit,shadows,calc}

\usepackage{color}

\usepackage[draft,margin]{fixme} %
\fxsetup{theme=color,mode=multiuser,status=final}
\FXRegisterAuthor{JL}{aJL}{JL}
\FXRegisterAuthor{OP}{aOP}{OP}
\FXRegisterAuthor{DOK}{aDOK}{DOK}
\usepackage{listings}
\usepackage{float}
\usepackage{listings-rust}
\newcommand{\rustmc}{\textsc{RustMC}\xspace}
\newcommand{\genmc}{\textsc{GenMC}\xspace}
\newcommand{\mixer}{\textsc{Mixer}\xspace}
\newcommand{\loomignore}{{\color{red}{$\dagger$}}}

\makeatletter
\RequirePackage[bookmarks,unicode,colorlinks=true]{hyperref}%
   \def\@citecolor{blue}%
   \def\@urlcolor{blue}%
   \def\@linkcolor{blue}%

\def\orcidID#1{\smash{\href{http://orcid.org/#1}{\protect\raisebox{-1.25pt}{\protect\includegraphics{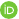}}}}}
\makeatother

\usepackage[firstpage,pos={0.99\paperwidth,0.5in},hanchor=r]{draftwatermark}
\SetWatermarkAngle{0}
\SetWatermarkText{\includegraphics[]{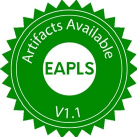}\includegraphics[]{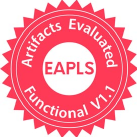}\includegraphics[]{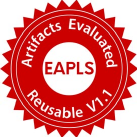}}
\begin{document}

\title{\rustmc{}: Automated Verification of Real-World Concurrent Rust
\thanks{Artefacts available in \url{https://doi.org/10.5281/zenodo.18806668}.}}

\author{Oliver Pearce\orcidID{0009-0004-1412-9961} \and
  Julien Lange\orcidID{0000-0001-9697-1378} \and
  Dan O'Keeffe\orcidID{0000-0003-3751-477X}}

\authorrunning{O.\ Pearce et al.}
\institute{Royal Holloway, University of London, UK}
\maketitle
\begin{abstract}
\input{abstract}
\end{abstract}

\input{intro}
\input{background}
\input{implementation_challenges}
\input{evaluation}

\input{related_work}
\input{conclusions}
\bibliographystyle{splncs04}
\bibliography{references}

\newpage
\appendix
\input{appendix}
\end{document}

%% file: abstract.tex
\rustmc is a stateless model checker that targets push-button 
verification of unmodified concurrent Rust programs.
As both Rust and C/C++ compile to LLVM IR, \rustmc builds on \genmc
which provides a verification framework for LLVM IR.
This allows verification of Rust code and any foreign function
dependencies. 
In this tool paper we present key implementation challenges, use-cases, and an
empirical evaluation of \rustmc.
The challenges include intercepting threading operations,
promoting memory intrinsics, and handling uninitialised
accesses, all stemming from Rust's unique compilation strategy.
Case studies adapted from real-world programs show \rustmc finds
concurrency bugs in unsafe Rust code, FFI calls to C/C++, and atomic
operations.
We evaluate our tool's functionality on the Loom
model checker test suite and a set of production-level Rust
concurrent data structures.

%% file: intro.tex
\section{Introduction}
Rust is a popular systems programming language that enforces memory and
thread safety through its ownership and borrowing system. Values in Rust have a
single owner and can be borrowed either mutably or immutably, with
mutable borrowing preventing any concurrent borrows of the same value.
These constraints effectively prevent data races in most safe Rust
code, as threads cannot make concurrent unsynchronised memory accesses
when any access is a write.
While Rust's type system 
prevents many of the common memory safety issues found in other systems 
programming languages, verification of concurrent programs remains an ongoing 
challenge. 
Concurrency bugs occur in Rust for a range of reasons, including
atomicity violations, ``unsafe'' Rust code and interactions between Rust
and C/C++ foreign function interface (FFI) dependencies.

In this tool paper, we extend
\genmc{}~\cite{Kokologiannakis21}, a stateless
model checker that systematically explores executions of concurrent C/C++
programs under weak memory models at the LLVM IR level, to support the
verification of multithreaded Rust programs.
By leveraging the LLVM compiler infrastructure's
intermediate representation (IR) we derive verifiable LLVM IR modules
from both Rust programs and their C/C++ dependencies.
\rustmc is, to the best of our knowledge, the only available framework
capable of exhaustively exploring the state space of Rust programs and
their dependencies, facilitating the detection of data races stemming
from Rust, C/C++ and interactions between the languages.
While our LLVM IR based approach allows us to verify mixed-language
programs, relying on this low-level representation introduces
challenges that impact on both \rustmc's out-of-the-box support for
real-world Rust code and the performance of
verification. Specifically, transforming an IR module emitted by the
Rust compiler into a verifiable representation requires significant
changes to \genmc{}'s compilation and transformation stages.
Below we describe how we addressed these challenges
and demonstrate the tool's applicability to real-world concurrent
programs.
The key contributions of this paper are:
\begin{myenumerate}
	\item We present \rustmc{}, an extension of \genmc{}, the first model
checker capable of verifying concurrent Rust programs and their FFI
dependencies;
\item We develop novel LLVM transformations to handle
	Rust-specific challenges, including LLVM memory intrinsics, mixed-size accesses and
uninitialised memory accesses, that arise from Rust's unique
compilation strategy;
\item We demonstrate \rustmc's effectiveness through examples
adapted from real-world programs, showing its ability to
detect data races in both pure Rust code and mixed-language programs
that use C/C++ dependencies;
\item We evaluate \rustmc's scalability and support for push-button
  verification of real-world Rust code. We evaluate our tool on two
  sets of examples. One consists of 72 tests adapted from the test
  suite of the Loom model checker~\cite{loom}, the other is based on
  tests from the 14 most popular crates providing concurrent data
  structures in Rust; and
\item We provide \rustmc{} as an open-source tool that extends
\genmc{} and enables the fully automated verification of concurrent Rust
applications.
Our artefact will consist of: the source code of \rustmc, its
documentation, and the datasets used in our evaluation. All are
readily available on GitHub. A video demonstrating the tool is also provided~\cite{rustmc}.
\end{myenumerate}

%% file: background.tex
\section{Background}\label{background}

\subsection{Concurrency Bugs in Rust}
While Rust code provides strong safety guarantees, concurrency bugs can still
occur. One potential source is seemingly safe Rust code
that harbours data races due to unsafe operations in foreign function
dependencies. For example, consider the simple Rust program shown in
Figure~\ref{fig:mixed_code_race_example}, which calls a C function
\texttt{inc\_C\_counter()} from multiple threads.
Rust's type system normally prevents concurrent access to shared
data. However, the C function bypasses these safeguards and directly
manipulates a global counter variable without synchronisation. Rust's
borrow checker cannot verify memory safety guarantees across language
boundaries.
\begin{figure}[t]
\begin{lstlisting}[language=Rust, numbers=left]
// Rust code
extern "C" {fn inc_C_counter();}
#[test]
fn test_C_counter() {
        thread::spawn(|| {unsafe {inc_C_counter();}});
        thread::spawn(|| {unsafe {inc_C_counter();}});
}
\end{lstlisting}
\begin{lstlisting}[language=C, numbers=left]
// C dependency
int counter = 0;
void inc_C_counter() { counter++; }
\end{lstlisting}
  \caption{A data race caused by an external, non-thread-safe, C function.}
  \label{fig:mixed_code_race_example}
\end{figure}
\rustmc successfully identifies the race condition automatically, detecting
a data race in which one thread attempts to read the \texttt{counter} while 
another thread simultaneously writes to it. Verification is performed
automatically and it is not required to rewrite code using a specific API, as is
the case with the Loom model checker~\cite{loom}.

Even safe Rust code can experience atomicity violations when using atomic types, 
as these types permit shared access and modification across multiple
threads~\cite{rust_reference_interior_mutability}.
Figure~\ref{fig:atomicity_violation} presents code adapted from an atomicity violation
bug reported in the \texttt{rand} crate~\cite{rand_crate} (also documented in a study of Rust
concurrent programming
bugs~\cite{yu2019fearlessconcurrencyunderstandingconcurrent}).
Function \texttt{is\_getrand\_available()} implements a common
pattern to cache the result of an expensive system call (\texttt{getrandom}) in a thread-safe way without
using a (more expensive) mutex.

\begin{figure}[t]  
\begin{lstlisting}[language=Rust, numbers=left, tabsize=1,
		style=boxed]
static CHECKED: AtomicI64 = AtomicI64::new(0);
static AVAILABLE: AtomicI64 = AtomicI64::new(0);
fn is_getrand_available() -> i64 {
    if (CHECKED.load(Ordering::Relaxed) == 0 ){ @\label{chk.load}@
        let mut buf: [u64; 0] = [];
        let result = getrandom(&mut buf);
        let available = if result == -1 {...} else { 1 };
        AVAILABLE.store(available, Ordering::Relaxed); @\label{av.store}@
        CHECKED.store(1, Ordering::Relaxed); @\label{chk.store}@
        available
    } else { AVAILABLE.load(Ordering::Relaxed) } @\label{av.load}@
}
fn main() {
    let t1 = thread::spawn(||{is_getrand_available()});
    let t2 = thread::spawn(||{is_getrand_available()});
    let r1 = t1.join().unwrap();
    let r2 = t2.join().unwrap();

    assert_eq!(r1, r2);  @\label{atom-assert}@
}  \end{lstlisting}

\caption{An atomicity violation in safe code, adapted from the Rand crate~\cite{rand_fix_commit}.}
\label{fig:atomicity_violation}
\end{figure}
Two \texttt{AtomicI64} variables are used to synchronise
operations:
\texttt{CHECKED} indicates whether \texttt{getrandom} has been
called before and \texttt{AVAILABLE} denotes the availability of
\texttt{getrandom};
both are first initialised to 0.
At compile time or run time, the atomic operations at
lines~\ref{av.store} and~\ref{chk.store} may be re-ordered such that one thread
loads \texttt{AVAILABLE} before it is set by another.

We show an example of an undesirable interleaving below, assuming
threads \texttt{t1} and \texttt{t2} are concurrently executing
\texttt{is\_getrand\_available()} in
Figure~\ref{fig:atomicity_violation}.
\begin{equation}\label{rand_interleaving}
  \centering
  \setlength{\tabcolsep}{12pt}
  \begin{tabular}{l@{\;}l||l@{\;}l}
    \multicolumn{2}{c}{\texttt{t1} } &   \multicolumn{2}{c}{\texttt{t2}}
    \\
    \hline
    \\ [-1em]
    {\texttt{CHECKED.store({\tiny\ldots})}}
                                     & {\small (line~\ref{chk.store})}
    \\[-0.5em]
                                     && 
                                        {\texttt{CHECKED.load({\tiny\ldots})}}
                                     &{\small(line~\ref{chk.load})}
    \\
                                     && {\texttt{AVAILABLE.load({\tiny\ldots})}}
                                     &{\small(line~\ref{av.load})}
    \\[-0.5em]
    {\texttt{AVAILABLE.store({\tiny\ldots})}}
                                     &{\small(line~\ref{av.store})}
  \end{tabular}
\end{equation}
Here, {\texttt{t1}} encountered \texttt{CHECKED==false}, and sets
both \texttt{CHECKED} and \texttt{AVAILABLE} to
\texttt{true}.
From {\texttt{t2}}'s point-of-view, if {\texttt{t1}}
has set \texttt{CHECKED} before setting \texttt{AVAILABLE},
then the function incorrectly returns false for {\texttt{t2}}.
To identify this bug, we must specify the property we are interested
in with an assert (line~\ref{atom-assert}). This assertion
characterises the consistency of concurrent calls to this function (i.e.,
all threads should see the same results, for all interleavings).
\rustmc can detect the bug and reports a trace,
see~\eqref{rand_interleaving}, violating atomicity.
The \texttt{unsafe} keyword disables the Rust 
compiler's safety checks to provide developers with greater flexibility. This
can lead to data races in Rust code without any FFI calls.
We show further real-world examples of concurrency bugs brought about by
the \texttt{unsafe} keyword in Appendix~\ref{app:examples} and 
online~\cite{rustmc}.
\subsection{\genmc{}: A Stateless Model Checker for C/C++} 
\genmc~\cite{Kokologiannakis21} supports
verification under several weak memory models (e.g., RC11, IMM, LKMM)
and represents program behaviours through execution graphs comprising
events (nodes) that correspond to individual memory accesses.
The tool's architecture consists of three main stages:
first, it compiles source code into LLVM IR (\eg using \texttt{clang});
second, it transforms this code to make it easier and faster to
analyse, e.g., bounding infinite loops, eliminating dead allocations,
etc.;
and finally, it conducts verification by exploring program
executions and checking for errors such as data races, assertion
violations, and memory safety issues.
Some of the latest development of the tool includes an optimal dynamic
partial order reduction (DPOR) algorithm that explores all possible
executions up to some equivalence relation while avoiding
redundant explorations~\cite{trust22}.
\genmc{} explores exactly one execution per equivalence class while
maintaining polynomial memory requirements.

\genmc\ provides an ideal foundation for extending stateless model
checking to Rust programs.
Its integration with LLVM's intermediate representation creates a
natural bridge for verifying Rust code, as both Rust and C/C++ compile
to LLVM IR.
This alignment enables \rustmc{} to verify not only pure Rust programs
but also their interactions with C/C++ dependencies.
Moreover, \genmc's three-stage architecture provides a modular framework that
\rustmc\ can extend to handle Rust-specific challenges.
As \genmc remains under active development, \rustmc stands to
benefit automatically from future enhancements to the underlying
verification framework without requiring significant modifications to
its Rust-specific components.
Notably, support for mixed-size accesses was recently added as part of
the \mixer{} extension to \genmc{}~\cite{kokolgiannakis_mixed_size}. As we
discuss in the next section, mixed-size access support is essential
for push-button verification of many real-world Rust programs.

%% file: implementation_challenges.tex
\section{\rustmc{}: Overview and Implementation
Challenges}\label{implementation_challenges}

We next present \rustmc{}, an extension to the \genmc{} framework for
verifying Rust applications. Figure~\ref{fig:arch-rustmc} gives an
overview of the framework. As in \genmc{}, there are three
stages: compilation, transformation and verification. \rustmc{} also
includes an automated test harness component (not shown) that
enables a Rust crate's existing test suite to serve as input drivers
for verification.
\input{fig-diagram}

\rustmc{} first compiles the application to LLVM IR, then
the \texttt{llvm-link} tool~\cite{llvm_link} is used to link the IR
file with the
\genmc{} verification component. This allows for the interception and
tracking of relevant
threading events. If the application contains FFI calls to C/C++ code, the
C IR is also linked.
\rustmc{} transforms the resulting IR using a series of LLVM passes to
prepare it for verification. In comparison to \genmc{}, \rustmc{}
introduces additional passes that address implementation
challenges relating to uninitialised memory accesses
(see~\S\ref{sec:undef}), calls to external functions
(see~\S\ref{sub:std-inline}) and unsupported LLVM intrinsics
(see~\S\ref{sec:llvm_intrinsics}).

Finally, \rustmc{} feeds the transformed IR file to the verification component.
This executes the \genmc{} stateless model checking algorithm to
detect any data races or assertion violations across all potential
thread interleavings.

\subsection{Externally Linked Dependencies}\label{sub:std-inline}
To control a program's schedule, a model checker 
must intercept thread creation and synchronisation operations. 
In C/C++, these operations appear directly in a program's compiled IR, allowing
\genmc{} to redefine the corresponding symbols (\eg
\texttt{pthread\_\allowbreak{}create}) with its 
own internal wrapper functions (\eg \texttt{\_\_VERIFIER\_thread\_\allowbreak{}create}).
However, for Rust applications, many standard library functions are externally
linked. As a result, thread creation and synchronisation operations are not 
visible in the compiled IR.

A naive solution to this problem is to build the entire standard library
alongside the application to expose threading calls.
However, this approach has several downsides: first, the larger codebase
significantly increases the engineering effort required to handle edge cases
and unsupported LLVM behaviours; second, processing the much larger IR through
each LLVM transformation pass becomes significantly slower; and third, the
increased code complexity breaks LLVM optimisation passes that \genmc{} relies
on.

Instead \rustmc{} leverages the \texttt{lto=fat} compilation flag to
perform link-time optimisation across the target crate and its dependencies, pulling standard library code into
standalone LLVM IR modules~\cite{fat_lto_rustc_book}. This targeted approach keeps the IR size
manageable and avoids most LLVM transformation issues.
One issue is that the standard library's thread creation implementation queries
environment variables and calls native threading internals to set a stack size for new threads. 
Since \genmc{}'s interpreter does not model environment variables or the
required native internals, calls to \texttt{thread::spawn} result in a segmentation fault.
To mitigate this issue we use an LLVM pass to preinitialise the stack size
of new threads to the default value for Rust on Linux.
\subsection{Handling LLVM Intrinsics}\label{sec:llvm_intrinsics}
LLVM supports a number of memory-related intrinsic functions such as those
corresponding to the C standard library functions \texttt{memcpy()},
\texttt{memmove()}, and \texttt{memset()}.
A call to one of these intrinsics may perform a series of
byte-wise accesses. For example, \texttt{memcpy(ptr \%dest, ptr \%src,
i64 24)} will copy 24 bytes from \texttt{\%src} to \texttt{\%dest}.
In \genmc{}, a custom LLVM pass promotes any intrinsic calls with
lengths evaluable at compile time into a series of typed
loads and stores. This allows the verification component to handle
them as individual accesses. Unpromotable memory intrinsics remain in
the IR and trigger a runtime error if they are executed during verification.
Rust's compilation strategy frequently produces intrinsics with lengths 
that cannot be resolved at compile time, necessitating a runtime fallback.

An IR-level implementation would require the use of integer types, as
LLVM does not have a representation for raw memory
values.
This introduces two semantic issues. First, integer types do not
carry provenance
information, so provenance is lost during copies.
This can cause alias analysis to miss pointer
escapes, potentially affecting any transformation passes that reason about
aliasing.
Second, manual implementations of \texttt{memcpy} may
over-propagate \texttt{poison}. For example, if a \texttt{memcpy} is
implemented using an \texttt{i64} load and some bits within the
\texttt{i64} are \texttt{poison}, the load returns poison for the
entire \texttt{i64}~\cite{mitenkov2021bytetyppe}.
By handling memory intrinsics inside the interpreter as direct byte-by-byte
copies, \rustmc{} avoids both problems.

This change is sound because the intrinsic implementations
are only invoked at runtime by the interpreter, after the
\texttt{IntrinsicLowering} pass has already converted intrinsic calls into
external function calls. Passes earlier in the pipeline may rely on
\texttt{@llvm.memcpy}'s intrinsic semantics, which guarantee that
the source and destination are either equal or non-overlapping. This
guarantee would be lost in any IR-level implementation since passes
would observe an ordinary function call with no non-overlapping constraint.
Note that \genmc{}'s custom memory intrinsic promotion pass also runs after the
\texttt{IntrinsicLowering} pass, preserving \texttt{memcpy} semantics for any
prior passes which rely on alias analysis.
The \mixer{}~\cite{kokolgiannakis_mixed_size} extension to \genmc{} is used to
support any subsequent non-bytewise accesses to memory written bytewise by the
interpreter fallback.
To reduce the performance overhead associated with mixed-size
accesses we run the static promotion pass first, transforming as many
intrinsics as possible into loads and stores which match the size of
subsequent accesses.

In addition to memory intrinsics, we encountered calls to the
\texttt{sse2.pause} intrinsic for optimising the performance of spin loops.
As this intrinsic is not supported by \genmc{}'s interpreter and does
not affect verification we introduce a pass to remove any calls to it.
\subsection{Aggregate Types}\label{sec:constant_aggregates}
The Rust compiler emits many values of LLVM aggregate type (\eg
structs, arrays) and frequently uses them in patterns not 
supported by the verification component, presenting two problems.

First, the compiler can emit statically
evaluable \texttt{extractvalue} instructions to extract an element
from an aggregate. However, the interpreter 
does not support the direct conversion of constant aggregates
into its internal value representation, which is common
in IR derived from Rust programs. We transform these
\texttt{extractvalue} instructions into constants with a custom
\texttt{FoldInterpreter\allowbreak{} UnsupportedConstants} pass based on LLVM's
\texttt{ConstantFoldExtractValueIns\allowbreak{} truction()} function.
Figure~\ref{fig:constant_fold_example} illustrates an example of this
transformation
for LLVM IR that extracts the second field  from a constant struct with two
\texttt{i64} fields. The pass transforms the IR to a simple store of the
constant field value (\texttt{i64 10}).

\begin{figure}[t]
  \begin{lstlisting}[language=LLVM, numbers=left,  escapechar=|]
%1 = extractvalue { i64, i64 } { i64 0, i64 10 }, 1
store i64 %1, ptr %6, align 8
\end{lstlisting}
  \begin{lstlisting}[language=LLVM, numbers=left,  escapechar=|]
store i64 10, ptr %6
\end{lstlisting}
  \caption{FoldInterpreterUnsupportedConstants transformation of LLVM
    IR to extract elements from a constant struct (top) to a simple
  store (bottom).}
  \label{fig:constant_fold_example}
\end{figure}
Second, while shared globals are initialised using native LLVM functions, LLVM
delegates the initialisation of thread-local globals to the client application. \genmc{}
stores per-thread copies of thread-local globals in a dedicated hashmap and
initialises them using the LLVM \texttt{getConstantValue()} function, which does not support
aggregate types with concrete initialisers. The Rust compiler frequently emits thread-local 
globals of aggregate type, causing the interpreter to crash when initialising them.
\rustmc{} addresses this by decomposing thread-local globals into
their constituent bytes during initialisation, reconstructing them on access.
As only shared memory accesses are added to the event graph, this change does
not affect the execution graphs produced during verification.

\subsection{Undefined Values}\label{sub:undef}
\label{sec:undef}
LLVM supports undefined (\texttt{undef}) values to represent indeterminate
values. Each time an \texttt{undef} value is used by an instruction a
different bit pattern may be observed.
Typically, \texttt{undef} models loading from uninitialised
memory~\cite{LLVM_LOAD_SEMANTICS,taming_undefined_behaviour}. This is permitted
as long as the \texttt{undef} value is not used in a subsequent operation that
may trigger immediate undefined behaviour for some potential bit
pattern~\cite{LLVM_UB_MANUAL}.
For example, dividing any value by an \texttt{undef} value leads to undefined
behaviour as the \texttt{undef} value could be 0.
\genmc{}'s model of memory accesses does not handle uninitialised
loads (it requires that each read has a
preceding write to take its value from).
As loads from uninitialised memory have no preceding write the tool is
unable to construct an execution graph.

\begin{figure}[h]
  \begin{lstlisting}[language=Rust, numbers=left]
pub fn foo() -> Result<(), i64> { Ok(()) }
\end{lstlisting}

  \begin{lstlisting}[language=LLVM, numbers=left,  escapechar=|]
define { i64, i64 } @foo() unnamed_addr #0 !dbg !7 {
  %_0 = alloca [16 x i8], align 8 ; allocate 16 bytes
  store i64 0, ptr %_0, align 8, !dbg !35 ; put 0 in 1st chunk
  %0 = load i64, ptr %_0, align 8, !dbg !36 ; initialised read
  %1 = getelementptr inbounds i8, ptr %_0, i64 8, !dbg !36
  %2 = load i64, ptr %1, align 8, !dbg !36 ; uninitialised read |\label{lst:undef-load}|
  %3 = insertvalue { i64, i64 } poison, i64 %0, 0, !dbg !36
  %4 = insertvalue { i64, i64 } %3, i64 %2, 1, !dbg !36
  ret { i64, i64 } %4, !dbg !36  |\label{lst:undef-ret}|
}
\end{lstlisting}
  \caption{Uninitialised read in Rust (top) and corresponding LLVM IR (bottom).}
  \label{fig:unit_uninit_read}
\end{figure}

Figure~\ref{fig:unit_uninit_read} shows an uninitialised read arising from typical Rust
code. The function \texttt{foo()} returns the \texttt{Ok(T)} variant
of \texttt{Result<T, E>}.
A \texttt{Result<T, E>} is stored in memory as a
tagged union whose size is determined
by its largest variant.
When \texttt{T} is the zero-sized unit type \texttt{()}, the \texttt{Ok()}
variant occupies less space than is allocated for the union. The remaining bytes
are never written, yet must be loaded when \texttt{Result<(),i64>} is read.
This occurs at line~\ref{lst:undef-load} in Figure~\ref{fig:unit_uninit_read}. Nothing
is stored in the second 8-byte chunk of allocated memory, so the load yields
\texttt{undef}. The function returns \texttt{\{0, undef\}} where \texttt{0}
is the discriminant for the \texttt{Ok()} variant.

To handle uninitialised loads while meeting \rustmc{}'s verification
requirements, we implement an LLVM pass that explicitly
writes undefined values into each stack allocation.
This approach allows us to verify programs containing uninitialised
reads as \rustmc{} interprets an explicit ``\texttt{store undef}'' as
storing 0 to a memory location. Consequently each uninitialised read
has a preceding write, as required. Interpreting accesses to uninitialised
memory as 0 preserves program semantics provided there is no immediate
undefined behaviour arising from uninitialised values taking non-zero
bit patterns~\cite{LLVM_UB_MANUAL}.
We consider this an acceptable trade-off for concurrency verification
given the tool's focus on
detecting data races and memory ordering issues rather than detecting
all forms of undefined behaviour. Alternatives, such as exploring all
possible \texttt{undef} values, would cause state explosion and make
the verification of real-world code impractical.
We note that \rustmc{}'s undefined value
transformation currently only supports stack allocations. Extending it to
handle undefined values in dynamic heap allocations using a similar
approach should be possible, but is currently unsupported by the tool.

\subsection{Soundness, completeness and optimality}
As \genmc's soundness, completeness and optimality guarantees depend on 
preserving the semantics of a program under test, the LLVM transformations 
introduced by \rustmc{} are designed to be semantics-preserving.
Our undefined value handling explores all execution traces in which \texttt{undef}
is consistently observed as 0 in programs that misuse uninitialised
memory. For all other programs, all possible execution traces described
by the LLVM semantics are explored.

While promoting memory intrinsics into sequences of typed loads and
stores and handling any non-constant intrinsics with a custom internal
implementation can increase the number of events in execution graphs --- and
consequently the number of potential interleavings --- this does not
compromise \genmc's optimality. The theoretical properties of \genmc's
verification algorithm~\cite{trust22} are defined at the level of
execution graphs and their exploration; RustMC inherits these
properties directly as it reuses \genmc's verification component
without modification. Thus, \rustmc{} maintains the same formal
guarantees as \genmc while extending support to Rust-specific
compilation patterns.

We note that \rustmc{} employs the \mixer{} algorithm of \genmc{} to
handle mixed-size accesses. Although \mixer{} is parametric in its
choice of memory model, the current implementation is
restricted to the Release-Acquire (RA) fragment of RC11.

%% file: fig-diagram.tex
\begin{figure}[htb]\centering
\begin{tikzpicture}[
    node distance = 2.5cm,
    task/.style = {
        rectangle,
        minimum width=3.0cm,
        minimum height=3.4cm,
        text centered,
        draw=black,
        fill=green!10,
        rounded corners,
        drop shadow
      },
     tasklabel/.style = {
        font=\bfseries
      },
    subtask/.style = {
        rectangle,
        minimum width=1.5cm,
        minimum height=0.4cm,
        text centered,
        draw=black,
        fill=white,
        rounded corners,
        font=\small
    },
    arrow/.style = {
        thick,
        ->,
        >=stealth
      },
      scale=0.7, every node/.style={transform shape}
]

\def\documentpath#1#2{
    \path (#1) coordinate (doc_center);
    \path (doc_center) +(-0.7,-0.5) coordinate (bottom_left)
                      +(0.7,-0.5) coordinate (bottom_right)
                      +(-0.7,0.5) coordinate (top_left)
                      +(0.7,0.1) coordinate (top_right)
                      +(0.4,0.5) coordinate (fold_start)
                      +(0.7,0.1) coordinate (fold_end);
    \fill[fill=yellow!10, drop shadow] 
        (top_left) -- (fold_start) -- (fold_end) -- 
        (top_right) -- (bottom_right) -- (bottom_left) -- cycle;
    \draw (top_left) -- (fold_start) -- (fold_end) -- 
          (top_right) -- (bottom_right) -- (bottom_left) -- cycle;
    \draw (fold_start) -- (fold_end);
    \node[font=\small, align=center] at (doc_center) {{\sf #2}};
}

\path (0,0) coordinate (doc0);
\documentpath{doc0}{\texttt{.rs}}

\path (0,-1.8) coordinate (cpp);
\documentpath{cpp}{\texttt{.c,.cpp}}

\node[draw,rectangle, minimum size=1.5cm,rounded corners,fill=yellow!10,drop shadow] at (3,-3.0) (stdlib) {
  \begin{tabular}{l}
    \texttt{std::sync},
    \\
    \texttt{std::thread},
    \\
    \texttt{std::sys}, \ldots
  \end{tabular}};

\node[task,minimum height=3.4cm, minimum width=3cm] at (3,0) (task1) {};
\node[subtask] (subtask13) at ([yshift=-0.6cm]task1)
	{\texttt{\texttt{\href{https://doc.rust-lang.org/rustc/codegen-options/index.html\#lto}{lto=fat}}}};
\node[subtask] (subtask14) at ([yshift=0.2cm]task1)
	{\texttt{\texttt{\href{https://llvm.org/docs/CommandGuide/llvm-link.html}{llvm-link}}}};

\node[task, right=of task1,minimum height=3.4cm, minimum width=3.8cm] (task2) {};
\node[subtask] (subtask4) at ([yshift=-1.3cm]task2) {\texttt{Undef}  (\S~\ref{sub:undef})};
\node[subtask] (subtask2) at ([yshift=0.8cm]task2) {\texttt{InitStackSize}(\S~\ref{sub:std-inline})};
\node[subtask] (subtask3) at ([yshift=-0.6cm]task2) {\texttt{FoldConstants}
(\S~\ref{sec:constant_aggregates})};
\node[subtask] (subtask5) at ([yshift=0.1cm]task2) {\texttt{PromotePause}  (\S~\ref{sec:llvm_intrinsics})};

\node[task, right=of task2] (task3) {};

\node[tasklabel] at ([yshift=-0.3cm]task1.north) {Compilation};
\node[tasklabel] at ([yshift=-0.3cm]task2.north) {Transformation};
\node[tasklabel] at ([yshift=-0.3cm]task3.north) {Verification};

\node at (task3) {\includegraphics[width=.19\textwidth]{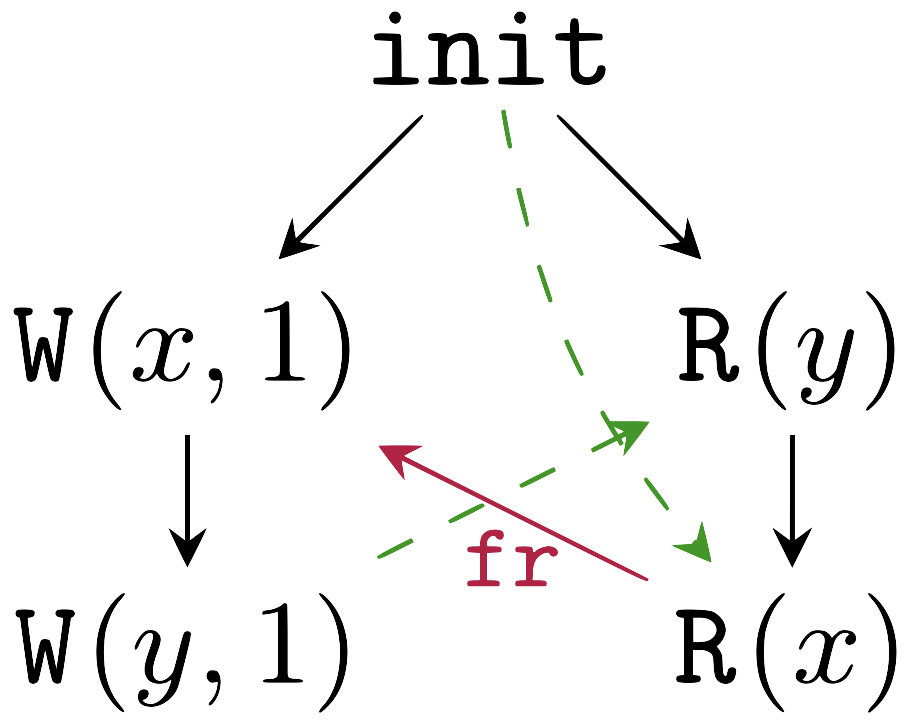}};
\node[tasklabel,font=\small] at ([yshift=0.3cm]task3.south) {(as \genmc's)};

\draw[arrow] (stdlib) -- (task1);
\draw[arrow] ([xshift=0.7cm]doc0) -- (task1.west);
\draw[arrow,densely dotted] ([yshift=0.52cm]cpp) -- ([yshift=-0.52cm]doc0.south);
\draw[arrow] (task1) -- node[midway] (doc1) {} (task2);
\documentpath{doc1}{\begin{tabular}{c}
                      LLVM \\ IR
                    \end{tabular}
                  }

\draw[arrow] (task2) -- node[midway] (doc2) {} (task3);
\documentpath{doc2}{\begin{tabular}{c}
                      LLVM \\ IR
                    \end{tabular}
                  }
                    
\end{tikzpicture}
\caption{Architecture of \rustmc.}\label{fig:arch-rustmc}
\end{figure}

%% file: evaluation.tex
\section{Evaluation}
We evaluate \rustmc{} through two complementary experiments that
measure its correctness on established benchmarks and its
applicability to production-level concurrent data structures.

The first experiment assesses our tool's bug-detection accuracy and
practical efficiency on 72 concurrent tests~\cite{loom-tests} adapted
from the Loom model checker test suite.
The second experiment evaluates \rustmc{}'s support
for real-world Rust language features by verifying test functions from
14 popular concurrency-focused crates.

Both experiments use functions marked with the \texttt{\#[test]} attribute
as self-contained verification harnesses. For each test we explore all concurrent behaviours and consider
verification successful if \rustmc{}'s output matches the expected behaviour.
A test's expected behaviour is considered to be a panic when annotated with 
the \texttt{\#[should\_panic]} attribute, or when its body
exercises Rust's unwinding functionality. Because \rustmc{} uses the
\texttt{panic=abort} strategy, tests that rely on unwinding to recover from a
panic will instead abort. We accept a panic outcome as a successful verification for these tests.

\input{rq-loom}

\input{rq-crate}

%% file: rq-loom.tex
\subsection{Experiment 1: \rustmc\ and the Loom Test Suite}\label{Experiment_1}

To evaluate \rustmc's effectiveness on established benchmarks, we
assessed its performance on 72 concurrent Rust tests adapted from the
Loom model checker test suite~\cite{loom-tests}. These tests cover a
range of concurrency primitives including atomics, mutexes, read-write
locks, condition variables, thread-local storage, and
\texttt{UnsafeCell<T>} (primitive for interior mutability in Rust)
operations.
This constitutes a demanding evaluation: the tests were designed for a
production-level model checker operating at the Rust API level, while
our research tool works at the lower LLVM IR abstraction, where
the same patterns manifest with significantly greater implementation
complexity.

For each test, we record whether \rustmc's outcome (success, panic,
or crash) matches the expected behaviour. We also measure
resource consumption: the number of executions explored, 
maximum memory usage, and CPU time (broken down
into the transformation and verification phases).
This evaluation aims to measure \rustmc's bug-detection accuracy (false
negatives and false positives) and its practical efficiency on
real-world concurrency scenarios.

\input{tbl-loom-tests}

Loom requires tests to be written using the Loom API, hence they
require adaptation to be used by a push-button model checker like
\rustmc.
Notably if one wants to model check program
\lstinline[language=Rust]!P! with Loom, then one must wrap it around
the construct
\lstinline[language=Rust]!loom::model(|| { P })!
which will tell Loom to explore all possible interleavings of
\lstinline[language=Rust]!P!.
All tests used in this experiment require this wrapper to be removed.
In addition, some Loom-specific calls must be re-interpreted (see
comment on \texttt{spec} below).

Out of the 20 test files in the Loom repository, we rule out three as
they either focus on higher-level concurrency issues or primitives we
do not target (e.g., channels, memory leak, deadlocks) or aim at
testing the Loom API itself.
In the remaining 17 files, shown in the first column of
Table~\ref{tab:results}, we additionally omit three further tests
aimed at testing the Loom API.
For all files, we run \rustmc{} with the \genmc{} flag
\texttt{--unroll=1} except for \crate{smoke} where it is necessary to
unroll to 3 to detect the bug (highlighted with {\color{red} $\ast$}).

Table~\ref{tab:results} gives the results of our experiment.
Out of the 72 tests, our tool currently supports 66 tests, it crashes
for the remaining 6 ($\approx 9.0\%$).
Out of these 66 supported tests, 60 ($\approx 91.0\%$) returned
the expected outcome.
Out of the 6 misdiagnosed tests, two are not supported by \mixer{}'s
memory model, nor by Loom, i.e., Loom returns the wrong result too
(they are flagged as \texttt{\#[ignore]}) because it does not fully
model \texttt{Ordering::Relaxed}.
Similarly, the four unexpected panics in \texttt{fences} are due to
sequentially-consistent fences which we do not support due to the current
limitation of \mixer{} to the RA memory model.
The 6 crashes are due to an external function with variable arguments (i.e.,
\texttt{syscall}).
Resource consumption is reasonable, with an average memory footprint
of 162.35 MB and mean CPU time of 8.75 seconds per test, though two
tests in the \texttt{spec} file require significantly more resources
(178 executions, ~233 MB, ~240s). These tests involve 4 threads looping
on a \texttt{load()} and invoking \texttt{thread::yield\_now()} in the
body of the loop.  The original tests include a Loom preemption bound of one (i.e.,
\texttt{builder.preemption\_bound = Some(1)}) which we approximated by
bounding the loop with one.

%% file: tbl-loom-tests.tex
\newcommand{\test}[1]{\texttt{#1}}
\newcommand{\crate}[1]{\texttt{#1}}

\begin{table}[p]
\centering
\scriptsize
\setlength{\tabcolsep}{2pt}
\renewcommand{\arraystretch}{0.65}
\caption{\rustmc{} verification results on tests adapted from Loom. Status:
	\textbf{S}: success, \textbf{P}: panic, \textbf{C}: crash. Expected:
	\textcolor{green!60!black}{$\checkmark$}: matches expected,
	\textcolor{red}{$\times$}: does not match. Executions: executions
	explored, Mem: max memory (MB), CPU: user time (s),
	P.Mem/P.CPU:
	Transformation pipeline max memory (MB) and user time (s).
  \loomignore{} indicates a test flagged with \texttt{\#[ignore]} in the Loom test suite.
}
\label{tab:results}
\resizebox{\textwidth}{!}{%
\begin{tabular}{@{}llcccrrrr@{}}
\toprule
\textbf{File} & \textbf{Test} & \textbf{Status} & \textbf{Expected} & \textbf{Executions} & \textbf{Mem} & \textbf{CPU} & \textbf{P.Mem} & \textbf{P.CPU} \\
\midrule
	\multirow{5}{*}{\crate{arc}}
	& \test{basic\_usage} & \textbf{S} & \textcolor{green!60!black}{$\checkmark$} & 4 & 155.6 & 0.86 & \multirow{5}{*}{179.9} & \multirow{5}{*}{2.09} \\
	& \test{sync\_in\_drop} & \textbf{S} & \textcolor{green!60!black}{$\checkmark$} & 3 & 153.0 & 0.83 & & \\
	& \test{try\_unwrap\_succeeds} & \textbf{S} & \textcolor{green!60!black}{$\checkmark$} & 1 & 145.8 & 0.51 & & \\
	& \test{try\_unwrap\_fails} & \textbf{S} & \textcolor{green!60!black}{$\checkmark$} & 1 & 145.8 & 0.41 & & \\
	& \test{try\_unwrap\_multithreaded} & \textbf{C} & -- & -- & 148.0 & 0.48 & & \\
\midrule
	\multirow{2}{*}{\crate{atomic}}
	& \test{compare\_and\_swap\_reads\_old}\loomignore{} & \textbf{S} & \textcolor{red}{$\times$} & 4 & 155.4 & 1.48 & \multirow{2}{*}{179.3} & \multirow{2}{*}{2.23} \\
	& \test{fetch\_add\_atomic} & \textbf{S} & \textcolor{green!60!black}{$\checkmark$} & 2 & 151.3 & 0.72 & & \\
\midrule
	\multirow{6}{*}{\crate{atomic\_int}}
	& \test{xor} & \textbf{S} & \textcolor{green!60!black}{$\checkmark$} & 1 & 136.2 & 0.47 & \multirow{6}{*}{169.8} & \multirow{6}{*}{1.93} \\
	& \test{max} & \textbf{S} & \textcolor{green!60!black}{$\checkmark$} & 1 & 136.6 & 0.37 & & \\
	& \test{min} & \textbf{S} & \textcolor{green!60!black}{$\checkmark$} & 1 & 136.9 & 0.47 & & \\
	& \test{compare\_exchange} & \textbf{S} & \textcolor{green!60!black}{$\checkmark$} & 1 & 136.6 & 0.38 & & \\
	& \test{compare\_exchange\_weak}\loomignore{} & \textbf{S} & \textcolor{green!60!black}{$\checkmark$} & 1 & 136.8 & 0.37 & & \\
	& \test{fetch\_update} & \textbf{S} & \textcolor{green!60!black}{$\checkmark$} & 1 & 137.0 & 0.34 & & \\
\midrule
	\multirow{4}{*}{\crate{atomic\_relaxed}}
	& \test{compare\_and\_swap} & \textbf{S} & \textcolor{green!60!black}{$\checkmark$} & 1 & 150.5 & 0.56 & \multirow{4}{*}{181.7} & \multirow{4}{*}{2.09} \\
	& \test{check\_ordering\_valid} & \textbf{S} & \textcolor{green!60!black}{$\checkmark$} & 5 & 157.6 & 1.09 & & \\
	& \test{check\_ordering\_invalid\_1} & \textbf{P} & \textcolor{green!60!black}{$\checkmark$} & 10 & 163.5 & 1.86 & & \\
	& \test{check\_ordering\_invalid\_2} & \textbf{P} & \textcolor{green!60!black}{$\checkmark$} & 10 & 163.0 & 1.85 & & \\
\midrule
	\multirow{2}{*}{\crate{condvar}}
	& \test{notify\_one} & \textbf{C} & -- & -- & 147.9 & 0.57 & \multirow{2}{*}{179.8} & \multirow{2}{*}{2.12} \\
	& \test{notify\_all} & \textbf{C} & -- & -- & 152.4 & 1.13 & & \\
\midrule
	\multirow{7}{*}{\crate{fence}}
	& \test{fence\_sw\_base} & \textbf{S} & \textcolor{green!60!black}{$\checkmark$} & 2 & 159.0 & 1.06 & \multirow{7}{*}{187.9} & \multirow{7}{*}{2.14} \\
	& \test{fence\_sw\_collapsed\_store} & \textbf{S} & \textcolor{green!60!black}{$\checkmark$} & 2 & 159.2 & 1.00 & & \\
	& \test{fence\_sw\_collapsed\_load} & \textbf{S} & \textcolor{green!60!black}{$\checkmark$} & 2 & 159.4 & 0.88 & & \\
	& \test{sb\_fences} & \textbf{P} & \textcolor{red}{$\times$} & 2 & 158.8 & 0.95 & & \\
	& \test{fence\_hazard\_pointer} & \textbf{P} & \textcolor{red}{$\times$} & 1 & 159.1 & 0.89 & & \\
	& \test{rwc\_syncs} & \textbf{P} & \textcolor{red}{$\times$} & 4 & 170.5 & 2.87 & & \\
	& \test{w\_rwc} & \textbf{P} & \textcolor{red}{$\times$} & 3 & 166.9 & 2.80 & & \\
\midrule
	\multirow{2}{*}{\crate{futures}}
	& \test{atomic\_waker\_valid} & \textbf{C} & -- & -- & 149.2 & 0.46 & \multirow{2}{*}{179.5} & \multirow{2}{*}{2.28} \\
	& \test{spurious\_poll} & \textbf{S} & \textcolor{green!60!black}{$\checkmark$} & 2 & 151.4 & 0.73 & & \\
\midrule
	\multirow{2}{*}{\crate{litmus}}
	& \test{load\_buffering}\loomignore{} & \textbf{S} & \textcolor{green!60!black}{$\checkmark$} & 3 & 150.8 & 0.86 & \multirow{2}{*}{178.0} & \multirow{2}{*}{2.05} \\
	& \test{store\_buffering} & \textbf{S} & \textcolor{green!60!black}{$\checkmark$} & 4 & 152.2 & 1.32 & & \\
\midrule
	\multirow{3}{*}{\crate{mutex}}
	& \test{mutex\_enforces\_mutual\_exclusion} & \textbf{S} & \textcolor{green!60!black}{$\checkmark$} & 1 & 151.5 & 0.57 & \multirow{3}{*}{182.0} & \multirow{3}{*}{2.31} \\
	& \test{mutex\_establishes\_seq\_cst} & \textbf{S} & \textcolor{green!60!black}{$\checkmark$} & 15 & 163.3 & 2.02 & & \\
	& \test{mutex\_into\_inner} & \textbf{S} & \textcolor{green!60!black}{$\checkmark$} & 1 & 151.6 & 0.76 & & \\
\midrule
	\multirow{10}{*}{\crate{rwlock}}
	& \test{rwlock\_read\_one} & \textbf{S} & \textcolor{green!60!black}{$\checkmark$} & 1 & 152.8 & 0.59 & \multirow{10}{*}{183.0} & \multirow{10}{*}{2.55} \\
	& \test{rwlock\_read\_two\_write\_one} & \textbf{S} & \textcolor{green!60!black}{$\checkmark$} & 2 & 153.2 & 0.65 & & \\
	& \test{rwlock\_write\_three} & \textbf{S} & \textcolor{green!60!black}{$\checkmark$} & 2 & 154.3 & 0.62 & & \\
	& \test{rwlock\_write\_then\_try\_write} & \textbf{S} & \textcolor{green!60!black}{$\checkmark$} & 1 & 148.4 & 0.39 & & \\
	& \test{rwlock\_write\_then\_try\_read} & \textbf{S} & \textcolor{green!60!black}{$\checkmark$} & 1 & 148.7 & 0.44 & & \\
	& \test{rwlock\_read\_then\_try\_write} & \textbf{S} & \textcolor{green!60!black}{$\checkmark$} & 1 & 148.9 & 0.46 & & \\
	& \test{rwlock\_try\_read} & \textbf{S} & \textcolor{green!60!black}{$\checkmark$} & 1 & 147.9 & 0.42 & & \\
	& \test{rwlock\_write} & \textbf{S} & \textcolor{green!60!black}{$\checkmark$} & 1 & 147.8 & 0.40 & & \\
	& \test{rwlock\_try\_write} & \textbf{S} & \textcolor{green!60!black}{$\checkmark$} & 1 & 148.0 & 0.42 & & \\
	& \test{rwlock\_into\_inner} & \textbf{S} & \textcolor{green!60!black}{$\checkmark$} & 1 & 151.8 & 0.56 & & \\
\midrule
	\crate{rwlock\_r1}
	& \test{rwlock\_two\_writers} & \textbf{S} & \textcolor{green!60!black}{$\checkmark$} & 24 & 176.2 & 11.43 & 179.2 & 2.09 \\
\midrule
	\crate{smoke}{\color{red} $\ast$}
	& \test{checks\_fail} & \textbf{P} & \textcolor{green!60!black}{$\checkmark$} & 0 & 170.0 & 5.74 & 179.8 & 2.22 \\
\midrule
	\multirow{3}{*}{\crate{spec}}
	& \test{relaxed}\loomignore{} & \textbf{S} & \textcolor{red}{$\times$} & 15 & 174.8 & 8.98 & \multirow{3}{*}{186.3} & \multirow{3}{*}{2.30} \\
	& \test{acq\_rel} & \textbf{S} & \textcolor{green!60!black}{$\checkmark$} & 178 & 232.9 & 239.68 & & \\
	& \test{test\_seq\_cst}\loomignore{} & \textbf{S} & \textcolor{green!60!black}{$\checkmark$} & 178 & 233.2 & 241.92 & & \\
\midrule
	\multirow{7}{*}{\crate{thread\_api}}
	& \test{initial\_thread} & \textbf{S} & \textcolor{green!60!black}{$\checkmark$} & 1 & 192.7 & 0.61 & \multirow{7}{*}{261.0} & \multirow{7}{*}{3.44} \\
	& \test{many\_joins} & \textbf{S} & \textcolor{green!60!black}{$\checkmark$} & 1 & 196.5 & 0.73 & & \\
	& \test{alt\_join} & \textbf{S} & \textcolor{green!60!black}{$\checkmark$} & 3 & 207.3 & 4.73 & & \\
	& \test{threads\_have\_unique\_ids} & \textbf{C} & -- & -- & 202.9 & 3.00 & & \\
	& \test{thread\_names} & \textbf{C} & -- & -- & 202.8 & 2.87 & & \\
	& \test{thread\_stack\_size} & \textbf{S} & \textcolor{green!60!black}{$\checkmark$} & 1 & 234.5 & 58.80 & & \\
	& \test{park\_unpark} & \textbf{S} & \textcolor{green!60!black}{$\checkmark$} & 1 & 192.7 & 0.67 & & \\
\midrule
	\multirow{3}{*}{\crate{thread\_local}}
	& \test{thread\_local\_test} & \textbf{S} & \textcolor{green!60!black}{$\checkmark$} & 1 & 152.1 & 1.00 & \multirow{3}{*}{178.9} & \multirow{3}{*}{2.12} \\
	& \test{nested\_with} & \textbf{S} & \textcolor{green!60!black}{$\checkmark$} & 1 & 143.9 & 0.36 & & \\
	& \test{drop\_test} & \textbf{S} & \textcolor{green!60!black}{$\checkmark$} & 1 & 151.6 & 1.03 & & \\
\midrule
	\multirow{13}{*}{\crate{unsafe\_cell}}
	& \test{atomic\_causality\_success} & \textbf{S} & \textcolor{green!60!black}{$\checkmark$} & 2 & 165.9 & 0.85 & \multirow{13}{*}{202.5} & \multirow{13}{*}{2.60} \\
	& \test{atomic\_causality\_fail} & \textbf{P} & \textcolor{green!60!black}{$\checkmark$} & 0 & 163.3 & 0.61 & & \\
	& \test{unsafe\_cell\_race\_mut\_mut\_1} & \textbf{P} & \textcolor{green!60!black}{$\checkmark$} & 0 & 163.5 & 0.61 & & \\
	& \test{unsafe\_cell\_race\_mut\_mut\_2} & \textbf{P} & \textcolor{green!60!black}{$\checkmark$} & 0 & 168.8 & 1.34 & & \\
	& \test{unsafe\_cell\_race\_mut\_immut\_1} & \textbf{P} & \textcolor{green!60!black}{$\checkmark$} & 0 & 163.8 & 0.65 & & \\
	& \test{unsafe\_cell\_race\_mut\_immut\_2} & \textbf{P} & \textcolor{green!60!black}{$\checkmark$} & 0 & 163.6 & 0.61 & & \\
	& \test{unsafe\_cell\_race\_mut\_immut\_3} & \textbf{P} & \textcolor{green!60!black}{$\checkmark$} & 0 & 168.3 & 0.95 & & \\
	& \test{unsafe\_cell\_race\_mut\_immut\_4} & \textbf{P} & \textcolor{green!60!black}{$\checkmark$} & 0 & 168.0 & 0.98 & & \\
	& \test{unsafe\_cell\_race\_mut\_immut\_5} & \textbf{P} & \textcolor{green!60!black}{$\checkmark$} & 0 & 168.9 & 0.99 & & \\
	& \test{unsafe\_cell\_ok\_1} & \textbf{S} & \textcolor{green!60!black}{$\checkmark$} & 1 & 163.8 & 0.62 & & \\
	& \test{unsafe\_cell\_ok\_2} & \textbf{S} & \textcolor{green!60!black}{$\checkmark$} & 1 & 163.8 & 0.70 & & \\
	& \test{unsafe\_cell\_ok\_3} & \textbf{S} & \textcolor{green!60!black}{$\checkmark$} & 1 & 167.0 & 0.90 & & \\
	& \test{unsafe\_cell\_access\_after\_sync} & \textbf{P} & \textcolor{green!60!black}{$\checkmark$} & 5 & 169.7 & 1.13 & & \\
\midrule
	\crate{yield}
	& \test{yield\_completes} & \textbf{S} & \textcolor{green!60!black}{$\checkmark$} & 4 & 152.2 & 0.76 & 175.3 & 2.02 \\
\bottomrule
\end{tabular}
}%
\end{table}

%% file: rq-crate.tex
\subsection{Experiment 2: \rustmc\ and Popular Crates}\label{Experiment_2}

To evaluate \rustmc's applicability to real-world concurrent Rust
code, we conducted a characterisation study on tests from 14 popular
crates in the Rust ecosystem. Our aim is to establish a baseline for
push-button verification of real-world Rust code and identify areas
for future improvement.

\begin{figure}[t]
  \includegraphics[width=0.99\textwidth]{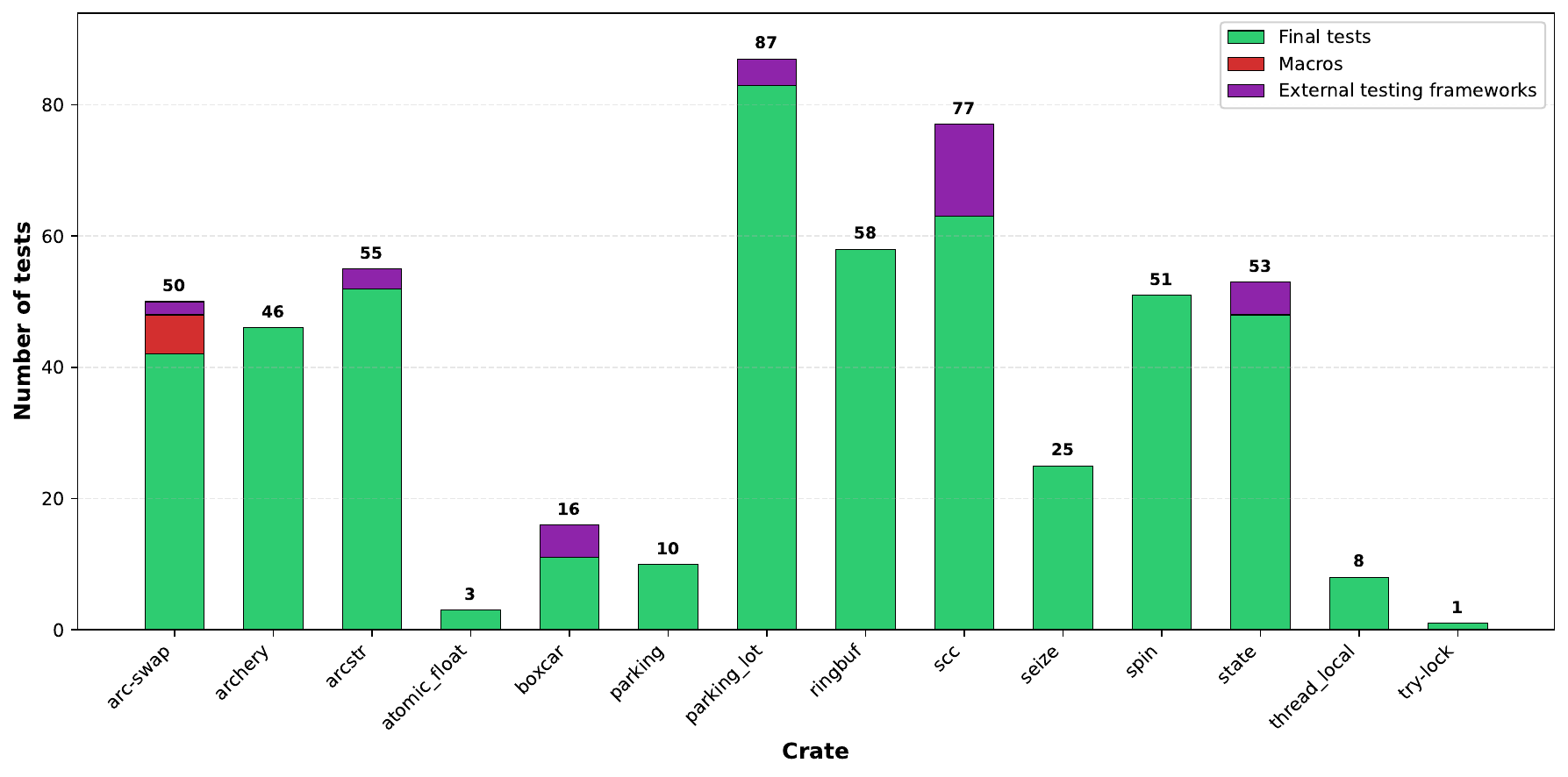}
  \caption{Selection of tests from popular concurrency-related
  crates.}\label{fig:rq-crate-selection}
\end{figure}

We selected crates by surveying those with the most
downloads in the last 90 days from the concurrency category of
\texttt{crates.io}.
Our current focus is on user-facing concurrent data structures, and so we excluded crates that serve primarily as internal building blocks or rely on higher-level concurrency models (e.g., asynchronous runtimes, channels). While asynchronous runtimes are supported by our tool, we leave a systematic evaluation of async crates to future work.
Figure~\ref{fig:rq-crate-selection} gives an overview of the number of
tests available for each crate.

\begin{figure}[!h]
  \includegraphics[width=0.99\textwidth]{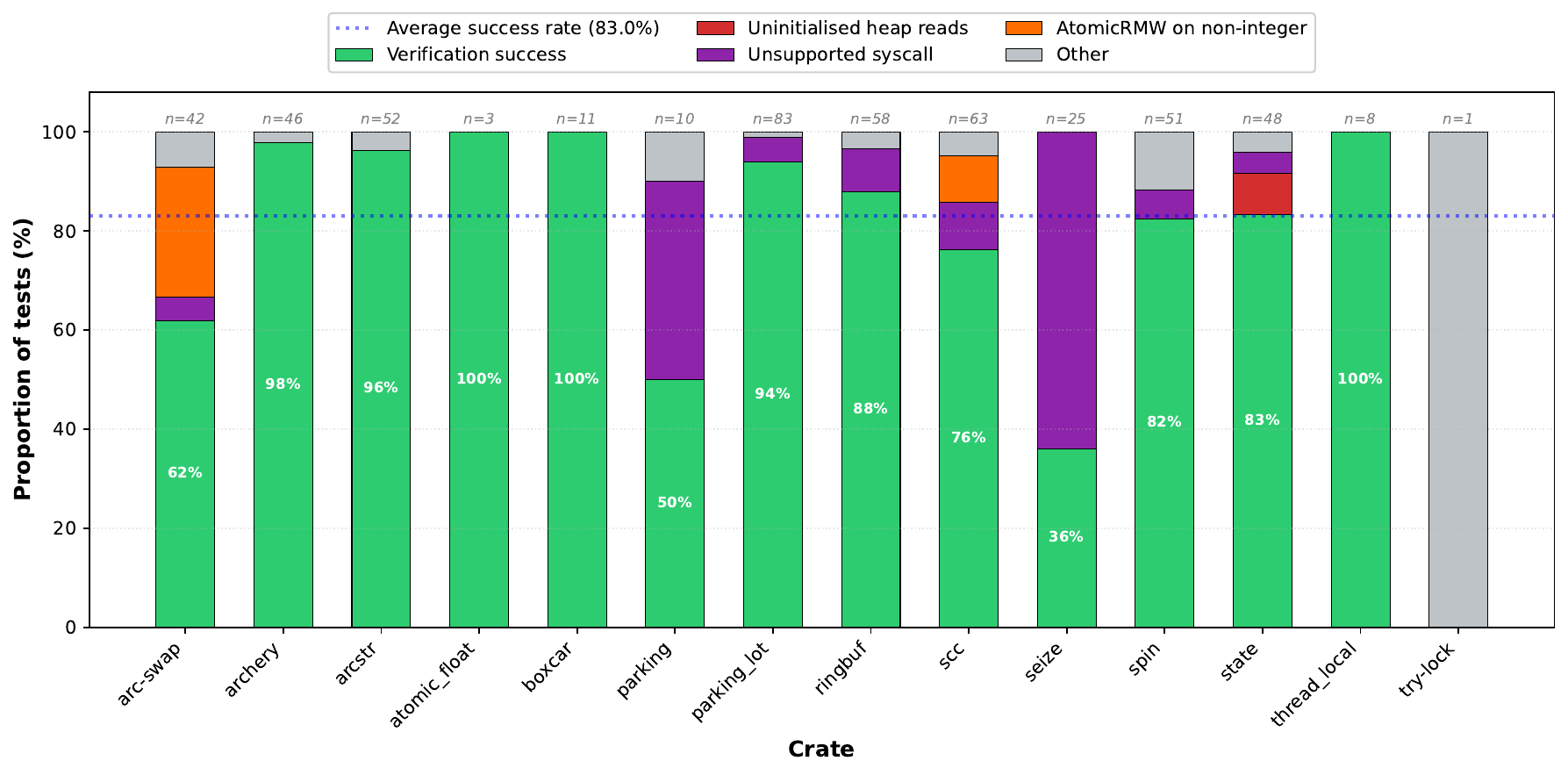}
  \caption{Results of model-checking tests from popular
  concurrency-related crates.}\label{fig:rq-crate-success}
\end{figure}

Our test-harness component automatically identifies all test
functions in a given crate and invokes \rustmc{} on them as targets for
verification.
Tests were filtered to remove those dependent on external testing
frameworks including Loom and Proptest (\textsf{External
testing frameworks} in \F\ref{fig:rq-crate-selection}). We also
excluded macro-generated test cases (\textsf{Macros} in
\F\ref{fig:rq-crate-selection}) as our test harness does not
currently support using them as entry points. 
This filtering process is conservative: for example, tests removed due to Loom
dependencies could potentially be adapted (as in Experiment 1), and
macro-generated tests represent an engineering challenge rather than a
fundamental limitation. All tests are run with the \texttt{--unroll=2} \genmc{} flag.
Figure~\ref{fig:rq-crate-success} presents verification results across
the 14 crates, with tests categorised by outcome.

\rustmc achieved an 83.0\% verification success rate over all
tests in our dataset. This includes tests from production-grade
concurrent data structures such as \texttt{archery} and
\texttt{boxcar}, demonstrating the tool's ability to scale from
synthetic benchmarks to complex, real-world concurrent Rust code.
Tests that fail to verify fall into three primary categories:
\begin{itemize}[leftmargin=*, itemindent=1em]
  \item \textsf{Uninitialised heap reads}: Uninitialised reads from
    dynamically allocated memory, e.g., when using Rust's
    \texttt{MaybeUninit} type, which requires extending our undefined
    value handling from stack to heap allocations.
  \item \textsf{Unimplemented syscalls}: External functions with variable
    arguments which we do not currently support.
  \item \textsf{AtomicRMW on non-integer}: Atomic operations on pointers, which are
not supported by the verification component.
\end{itemize}
The \textsf{Other} category in \F\ref{fig:rq-crate-success} notably
includes ($i$) a few unsupported intrinsics we conservatively
convert to runtime errors and ($ii$) four ($>$1h) timeouts due to unusually large
tests~\cite{spin-rsbiga,spin-rsbigb,spin-rsbigc,arc-swapbiga}.

The varying success rates across crates (from
near-complete verification in some to limited coverage in others)
reflects the diversity of concurrent programming patterns in the Rust
ecosystem.
Some crates heavily utilise language features that align well with
\rustmc{}'s current capabilities, while others employ patterns that push
the boundaries of our LLVM-level analysis.
The results provide a baseline for measuring future improvements and
identify specific areas (particularly dynamic memory patterns and
modelling common syscalls) where extending support would increase
coverage of real-world code.

%% file: related_work.tex
\section{Related Work}\label{sec:related-work}
Rust and its usage have been the focus of much research in
the last decade.
Notably, RustBelt~\cite{JKD18} provides formal verification of Rust's
type system and ownership model.
Evans et al.~\cite{EvansCS20} shows that while only 30\% of Rust
libraries explicitly use unsafe code, over half contain unsafe
operations somewhere in their dependency chain that bypass Rust's
static checks.
Astraukas et al.~\cite{AstrauskasMP0S20} shows that unsafe code
appears frequently, particularly for language interoperability, though
its unsafe code tends to be straightforward and well-contained.

Dynamic partial order reduction (DPOR) techniques have been
successfully applied to model check concurrent programs across
different languages and
contexts~\cite{AbdullaAJS14,FlanaganG05,AbdullaAAJLS15}. For Rust
specifically, several tools target concurrency bug detection, though
with different approaches and tradeoffs compared to ours.
Kani~\cite{kani} is a bounded model checker for Rust which identifies undefined
behaviour and verifies custom correctness properties provided by assertions or
function contracts for sequential programs. 

Miri~\cite{miri_paper} is part of the Rust compiler project and is
advertised as ``an undefined behavior detection tool for Rust''. It
works as an interpreter of Rust's mid-level IR.
Miri is particularly effective at detecting undefined behaviours, but
its inherently dynamic approach means that it cannot systematically
analyse all potential interleavings of concurrent programs.
Loom~\cite{loom} is an adaptation of CDSChecker~\cite{NorrisD16}. It
explores interleavings of concurrent code under
the C11 memory model using partial-order reduction techniques.
In comparison to \rustmc{}/\genmc{}, the CDSChecker algorithm is less
efficient since its backtracking algorithms can explore both redundant
and infeasible (prohibited by the memory model)
executions~\cite{weakly_libraries,trust22}.
Loom requires developers to port their code to use the Loom API,
as opposed to the push-button approach of \rustmc{}.
Shuttle~\cite{awshuttle} is another concurrent code testing tool for
Rust, developed at AWS. It is similar to (and inspired by) Loom but
aims at better scalability at the cost of soundness. Shuttle uses a
randomised scheduler~\cite{BurckhardtKMN10} to guarantee good
coverage.
Lockbud~\cite{Lockbud,QinCLZWSZ24} is a static bug detector for
blocking bugs and potential atomicity violations in Rust. It
identifies potential atomicity violations by detecting syntactical
patterns commonly found in atomicity violation bugs. While this
pattern-based approach can identify potential issues, \rustmc's
systematic exploration of thread interleavings allows it to
definitively verify atomicity properties through assert statements,
detecting actual violations rather than potential ones.
Other work~\cite{CRUST_TomanPT15,LiWSL22_FFI_checker} explores 
verification techniques for Rust's FFI dependencies, but 
focuses primarily on memory management bugs rather than concurrency
issues.

%% file: conclusions.tex
\section{Conclusions}\label{sec:conclusions}
\rustmc represents a significant step forward in verification of Rust code,
and opens new possibilities for detecting subtle concurrency bugs in
concurrent Rust programs and their FFI dependencies.
Extending \genmc to handle Rust's compilation characteristics
presented several technical challenges but its modular architecture
allowed us to re-use the back-end verifier.
Hence \rustmc will automatically benefit from future
improvements to \genmc's verification engine.

Looking ahead, \rustmc provides a foundation for systematic analysis
of real-world concurrent Rust programs, helping developers ensure
their code is free from concurrency issues.
Our future work will focus on applying \rustmc to verify more concurrent
code in Rust projects and their dependencies.

%% file: appendix.tex
\section{Additional examples}\label{app:examples}
\subsection{Out of bounds accesses}
Data races in Rust may also cause out of bounds accesses to allocated
memory, resulting in an immediate unrecoverable \texttt{panic!}
error~\cite{rust_panic_documentation}.
Figure \ref{fig:rustonomicon-data-race} gives an example of such a
bug.

\begin{figure}[htb]
\begin{lstlisting}[language=Rust, numbers=left, tabsize=1,
    style=boxed]
fn main() {
    let data = vec![1, 2, 3, 4];
    let idx = Arc::new(AtomicUsize::new(0));
    let other_idx = idx.clone();

    thread::spawn(move || { @\label{other_idx.ownership_move}@
        other_idx.fetch_add(10, Ordering::SeqCst); @\label{other_idx.increment}@
    });

    if idx.load(Ordering::SeqCst) < data.len() { @\label{idx.load}@
        unsafe {
            let i = idx.load(Ordering::SeqCst);
            let x = *data.get_unchecked(i); @\label{data.get_unchecked}@
        }
    }
}
\end{lstlisting}
  \caption{A data race which leads to an unrecoverable
  \texttt{panic!} error, adapted from~\cite{rustonomicon}.}
  \label{fig:rustonomicon-data-race}
\end{figure}
Variable \texttt{other\_idx} is a clone of the \texttt{Arc} type
used as a thread safe reference counter, i.e., \texttt{idx} and
\texttt{other\_idx} both contain references to the same value.
Recall that atomics have interior mutability and are not subjected to
restrictions on mutable aliasing.
An ownership move is performed at line~\ref{other_idx.ownership_move},
and the thread which has been passed ownership of
\texttt{other\_idx} increments the variable by 10.
At line~\ref{idx.load}, a bounds check is performed and if
\texttt{idx} is smaller than the length of the allocated
\texttt{vec}, then the position of \texttt{idx} in the allocated
\texttt{vec} array is accessed.
In an interleaving where \texttt{other\_idx} is incremented after
the bounds check at line \ref{idx.load}, an out-of-bounds read occurs
at line~\ref{data.get_unchecked}.\footnote{ Method
  \texttt{get\_unchecked()} is an unsafe method that retrieves an
  element from a vector without performing any bounds checking, i.e.,
  it is a potentially faster version of the regular \texttt{get()}
method.}
This leads to an immediate \texttt{panic!} call which is detected by
\rustmc along with the interleaving responsible for the error.
\subsection{Raw Pointers}
In Rust, developers can use raw pointers to bypass the language's
safety guarantees around aliasing and mutability when
necessary~\cite{rust_book_raw_pointers}.
In~\cite{QinCLZWSZ24}, Qin et al.\ show that sharing data between
threads through raw pointers to shared memory is a prevalent source of
concurrency bugs.
Figure~\ref{fig:raw-ptr-example} illustrates a concrete example of
such an issue.
\begin{figure}[htb]
\begin{lstlisting}
fn main() {
    let mut x: i64 = 2;
    let x_ptr = &mut x as *mut i64; // Pointer #1 to x @\label{x_ptr.create}@

    unsafe {
      let x_ptr_2: &mut i64 = &mut *x_ptr; // Pointer #2 to x @\label{x_ptr2.create}@
      let handle = thread::spawn(move || {
            *x_ptr_2 = 5; @\label{x_ptr2.mutate}@
        });
      *x_ptr = 10; @\label{x_ptr.mutate}@
      handle.join().unwrap();
    }
    let y = x; @\label{y.create}@
}
\end{lstlisting}
  \caption{A data race caused by the use of raw pointers for sharing
  data between threads.}
  \label{fig:raw-ptr-example}
\end{figure}
While raw pointers can be created in safe code as is seen at line
\ref{x_ptr.create}, dereferencing a raw pointer is always considered
unsafe and must be marked as such.
At line \ref{x_ptr2.create}, \texttt{x\_ptr} is dereferenced,
creating a second mutable reference to the value held at
\texttt{x}. Ownership of this reference is transferred
to a new thread where the underlying value \texttt{x} is mutated.
This is performed concurrently with
another write to \texttt{x} at line~\ref{x_ptr.mutate}.

Now \texttt{y} may be assigned different values (10 or 5)
depending on the observed interleaving. \rustmc is able to detect
this as a non-atomic race and reports the conflicting write events
at lines~\ref{x_ptr2.mutate} and~\ref{x_ptr.mutate}.

\subsection{Data Race in Thread-safe Traits}
We give another example that demonstrates how incorrect
implementations of Rust's thread safety traits can lead to data
races. The \texttt{Sync} trait, which marks types as safe to share
between threads, requires careful implementation to maintain Rust's
safety guarantees.
This type of bug has occurred in practice, e.g., the \texttt{internment}
crate contained a data race due to an unsafe implementation of
\texttt{Sync} for its \texttt{Intern<T>} struct~\cite{git_intern}. The
implementation allowed the struct to be shared between threads even
when instantiated with non-thread-safe types, potentially leading to
memory corruption.
To illustrate this class of bug, we create a minimal example where a
struct containing a \texttt{Cell<i64>} incorrectly implements
\texttt{Sync}. The \texttt{Cell} type provides interior
mutability but is explicitly designed to be non-thread-safe, making
this implementation unsound. Figure~\ref{unsafe_sync_example} shows
how this type definition can lead to data races.
\begin{figure}[t]
\begin{lstlisting}[language=Rust]
struct MyStruct { data: Cell<i64>, }

unsafe impl Send for MyStruct {} @\label{mystruct-send}@
unsafe impl Sync for MyStruct {} @\label{mystruct-sync}@

impl MyStruct {
    fn new(value: i64) -> Self {
        MyStruct { data: Cell::new(value), }
    }
    fn increment(&self) {self.data.set(self.data.get() + 1);}
}

fn main() {
    let foo = Arc::new(MyStruct::new(0)); @\label{mystruct-arc}@
    let foo_clone1 = foo.clone(); @\label{mystruct-clone1}@
    let foo_clone2 = foo.clone(); @\label{mystruct-clone2}@
    let t1 = thread::spawn(move || {foo_clone1.increment();});  @\label{t1.increment}@
    foo_clone2.increment(); @\label{main.increment}@
    t1.join().unwrap();
    let final_value = foo.data.get();
}
\end{lstlisting}
  \caption{Data race in safe Rust code.}\label{unsafe_sync_example}
\end{figure}

\texttt{MyStruct} contains a single field data which is a
\texttt{Cell<i64>}, i.e., a type that provides interior mutability
(data can be mutated even via an immutable reference).
Lines~\ref{mystruct-send} and~\ref{mystruct-sync} declare
\texttt{MyStruct} as safe to send between threads
(\texttt{Send}) and safe to share references across threads
(\texttt{Sync}). As they appear in an \texttt{unsafe} block, the
programmer is taking responsibility for thread safety.
However, \texttt{MyStruct} implements the \texttt{increment()} method,
which is clearly not thread-safe.
To expose the issue, we provide a \texttt{main()} function which
makes two concurrent calls to \texttt{increment()}.
Line~\ref{mystruct-arc} wraps a new \texttt{MyStruct} in an
\texttt{Arc} (Atomic Reference Counted), which allows safe sharing
between threads.
Lines~\ref{mystruct-clone1} and~\ref{mystruct-clone2} create new
references to \texttt{foo}; hence both threads are modifying the
same underlying data.
When given Figure~\ref{unsafe_sync_example} as input, \rustmc
correctly identifies the race. It detects a conflict between the
unsynchronised accesses to \texttt{foo} made by the main thread at
line \ref{main.increment} and the spawned thread at line
\ref{t1.increment}.